\documentclass[fleqn,10pt,bibtex]{wlscirep}
\usepackage{colortbl}
\usepackage{xcolor}
\usepackage{color}
\usepackage{amsmath}    
\usepackage{amssymb}
\usepackage{graphicx}   
\usepackage{float}
\usepackage[caption=false]{subfig}
\usepackage{stmaryrd}
\usepackage{tikz}

\title{Intrinsic noise and deviations from criticality in Boolean gene-regulatory networks}

\author[1]{Pablo Villegas}
\author[1,2]{Jos\'e Ruiz-Franco}
\author[1,3]{Jorge Hidalgo}
\author[1,*]{Miguel A. Mu\~noz}
 \affil[1]{Departamento de Electromagnetismo y F{\'i}sica de la Materia e Instituto Carlos I de F{\'i}sica Te{\'o}rica y Computacional. Universidad de Granada, E-18071 Granada, Spain}
 \affil[2]{Dipartimento di Fisica, Sapienza--Universit\`{a} di Roma, P.le A. Moro 5, 00185 Rome, Italy.}
 \affil[3]{Dipartimento di Fisica 'G.Galilei' and CNISM, INFN,Universit\'a di Padova, Via Marzolo 8, 35131 Padova, Italy.}

 \affil[*]{Correspondence and requests for materials should be addressed to M.A.M. mamunoz@onsager.ugr.es}
 
 \begin{abstract} Gene regulatory networks can be successfully modeled as Boolean networks. A much discussed hypothesis says that such model networks reproduce empirical findings the best if they are tuned to operate at criticality, i.e. at the borderline between their ordered and disordered phases. Critical networks have been argued to lead to a number of functional advantages such as maximal dynamical range, maximal sensitivity to environmental changes, as well as to an excellent trade off between stability and flexibility. Here, we study the effect of noise within the context of Boolean networks trained to learn complex tasks under supervision. We verify that quasi-critical networks are the ones learning in the fastest possible way --even for asynchronous updating rules-- and that the larger the task complexity the smaller the distance to criticality.  On the other hand, when additional sources of intrinsic noise in the network states and/or in its wiring pattern are introduced, the optimally performing networks become clearly subcritical. These results suggest that in order to compensate for inherent stochasticity, regulatory and other type of biological networks might become subcritical rather than being critical, all the most if the task to be performed has limited complexity.
   \end{abstract}

\begin{document}
  \flushbottom
 \maketitle 
  
  \section*{Introduction}
 The central dogma of molecular biology is that each single gene is transcribed into RNA, which in turn is translated into a protein, which --usually in cooperation with different proteins-- can regulate the expression of other genes, giving rise to a complex network of regulatory interactions and different possible patterns of gene expression \cite{Dogma}.  Genetic regulation, protein-protein interactions, as well as cell metabolic and signaling pathways are essential biological processes that can all be represented as networks \cite{buchanan2010}. The network picture encapsulates the complexity of cellular processes and provides us a natural framework for a systems-perspective approach to extremely complicated biological problems. As a matter of fact, the study of information processing in living systems has shifted from the analysis of single pathways to increasingly complex regulatory networks, allowing for a visualization of the collective effects of a host of units acting at unison.  Since the pioneering work of Kauffman \cite{kauffman69,kauffman1993,gros2011,dejong2002,alon2006}, genetic regulatory systems have been modeled as Boolean networks, in which the expression level of each gene is represented by a binary (on/off) variable and where mutual regulatory interactions are described as arbitrary random Boolean functions operating synchronously at discrete time steps. Even if admittedly simplistic and limited in a number of ways (e.g. continuous levels of gene expression might be essential to understand some cellular processes), such a binary description is particularly useful when dealing with large networks because it simplifies the overwhelming complexity of the real problem reducing it to a logical one. In particular, the Boolean approach has shed light on important conceptual problems such as the possibility of diverse (phenotypic) states emerging from a unique given genetic network, as well as the possibility of transitions among them (as happens in cell differentiation and reprogramming), and the emergence of cycles in cell states. The trajectory of the segment polarity network in the fly \textit{Drosophila melanogaster} \cite{droso} and the yeast cell cycle \cite{yeast} are two specific examples in which the most relevant features of gene expression have been fully elucidated on the basis of Boolean models \cite{bornholdt} (for more details we refer to the literature \cite{kauffman1993,gros2011,drossel2008,aldana-review,dejong2002}).

 Random Boolean networks (RBNs) can operate in different regimes including ordered and chaotic phases as well as a critical point (or line or surface) separating them in parameter space.  Ordered or frozen phases (typically obtained for small network connectivities) are characterized by a small set of stable attractors which are largely robust to perturbations, while in the disordered or chaotic phase (typically obtained for densely connected networks) perturbations rapidly propagate all through the network hindering the existence of truly stable states. As formalized mathematically by Derrida and Pomeau, separating these two phases there is a critical line (that used to be called the``edge of chaos'') at which perturbations propagate marginally \cite{derrida86}. It was conjectured some time ago that critical RBNs might be optimal to represent actual biological networks; the underlying idea is that operating at criticality might provide such systems with an optimal tradeoff between being exceedingly ordered/stable (thus, barely responsive to environmental changes, signals, and clues) and being too disordered/noisy (thus enormously sensitive to the effects of noise, lacking the required robustness and accuracy that biological machinery demands \cite{kauffman2003}). The criticality hypothesis states that the marginal situation between these two impractical  tendencies  constitutes an excellent compromise. 
This conjecture (which was developed in the machine-learning and neural-network community \cite{langton1990,maas2002,ber-Nat}), proposes that --by operating nearby criticality-- networks exhibit an optimal tradeoff between stability to perturbations and sensitivity/responsiveness to signals. Similarly, at larger timescales, it also provides an excellent compromise  between robustness and evolvability \cite{aldana2007,Aldana2012}. Moreover, it entails optimization of information storage and transmission \cite{ribeiro2008, basin}, response and sensitivity, computational capabilities, and a number of other functional advantages \cite{kauffman2003,aldana2007,Aldana2012,torres2012,Plenz-Functional,Kinouchi-Copelli,Kaneko2012,Physics,basin}.

In parallel, the development of powerful experimental high-throughput technologies in molecular biology has paved the way to experimental analyses of gene-expression patterns in large regulatory networks.  Recent empirical results, analyzing hundreds of microarray experiments to infer regulatory interactions among genes and implementing these data into Boolean models, seem to support the hypothesis that regulatory networks of \textit{Saccharomyces cerevisiae}, \textit{Escherichia coli}, \textit{Bacillus subtilis}, the murine macrophage, as well as some subnetworks of \textit{Drosophila melanogaster} and \textit{Arabidopsis thaliana} are indeed very close to criticality (in the sense of marginal propagation of perturbations) \cite{balleza2008, nykter2008}, while some other empirical analyses leave the door open to regulatory networks being ordered/subcritical \cite{shmulevich2005,kauffman2003}.

Recent work, aimed at rationalizing why and how criticality might come about in living systems, relies on adaptive/evolutionary models, in which communities of agents --each of them modeled as a Boolean network-- are selected for if they succeed at performing some complex tasks which may change in time. For instance, Hidalgo {\emph et al.} \cite{hidalgo2014} showed --by employing an information-theoretic approach-- that critical networks may emerge as optimal solutions in such a setting (however, the networks employed as a specific example in \cite{hidalgo2014} are fully connected and thus lack the structural richness of usual RBNs).  Similarly, Goudarzi {\emph et al.}  \cite{goudarzi2012} considered an ensemble of RBN's able to experience ``mutations'' in their topological structure and employed a genetic algorithm to select for those able to perform a given computational task (see Figure 1); i.e. networks which have learnt have a larger fitness than those that have not. Under these conditions the ensemble converges to a state in which all networks operate close to criticality.  In other words, critical networks emerge as the optimal solution out of the combined selective pressures of having to learn different tasks (i.e. having to produce different outcomes/attractors) and being able to readily shift among them following changes in the inputs in real time.

Given that living cells typically possess very low copy numbers of important regulatory molecules (e.g. for the $80\%$ of genes in \textit{Escherichia coli} genome the copy number of their associated proteins is less than $100$) \cite{gupta} stochastic effects are unavoidable and ubiquitous in gene regulatory networks \cite{noise2000}. Even if noise is usually assumed to be detrimental to reliable information transfer and, more in general, to cell functioning, stochastic effects can lead to beneficial outcomes; for instance, noise accounts for the observed (phenotypic) variability in identical (isogenic) populations \cite{elowitz2002} and can help cells to adapt to fluctuating environments \cite{elowitz2010,losick2008,balazsi2011,tkacik}.  Within the framework of RBN the role of stochasticity and noise has been addressed in a number of works \cite{stern,darabos,peixoto2012}.  

In this paper, we further delve in the problem of investigating the mechanisms and the conditions under which networks may become critical (or not), focusing on the role played by noise, and ask the question whether --in the presence of strongly noisy conditions-- regulatory networks, modeled as RBNs having to perform some complex computational task, operate in ordered, critical or supercritical regimes (see Figure 1). In other words: what is the role of noise in the emergence of criticality? Does it foster or hinder critical behavior?  In order to gauge the effect of noise on the dynamics of RBNs having to perform a complex task we consider a setting very similar to that of Goudarzi {\emph et al.}  \cite{goudarzi2012}, but including different additional sources of stochasticity. In particular, our approach differs from the previous one in three main aspects: (i) we consider asynchronous updating \cite{gershenson,drossel2005,peixoto-drossel2010} rather than the usual deterministic one, thus introducing the effect of stochasticity in the updating timings, (ii) both the structure and the dynamics of the networks are subjected to noise (be it intrinsic or external), and (iii) we do not consider an evolutionary algorithm to search for the best possible network connectivity, but rather we work in a constant-connectivity ensemble and explore how the network performance depends on the network connectivity, i.e. on the network dynamical state.

As we shall illustrate, criticality emerges as the solution providing the fastest route to learning complex tasks but, on the other hand, once additional sources of stochasticity are explicitly taken into account, ordered dynamical states perform better than critical ones. That is, networks need to compensate the excess of noise by becoming progressively more subcritical.

\section*{Model and training protocol} 

 \begin{figure*}[tb] 
 \centering \includegraphics[width=9cm,angle=0]{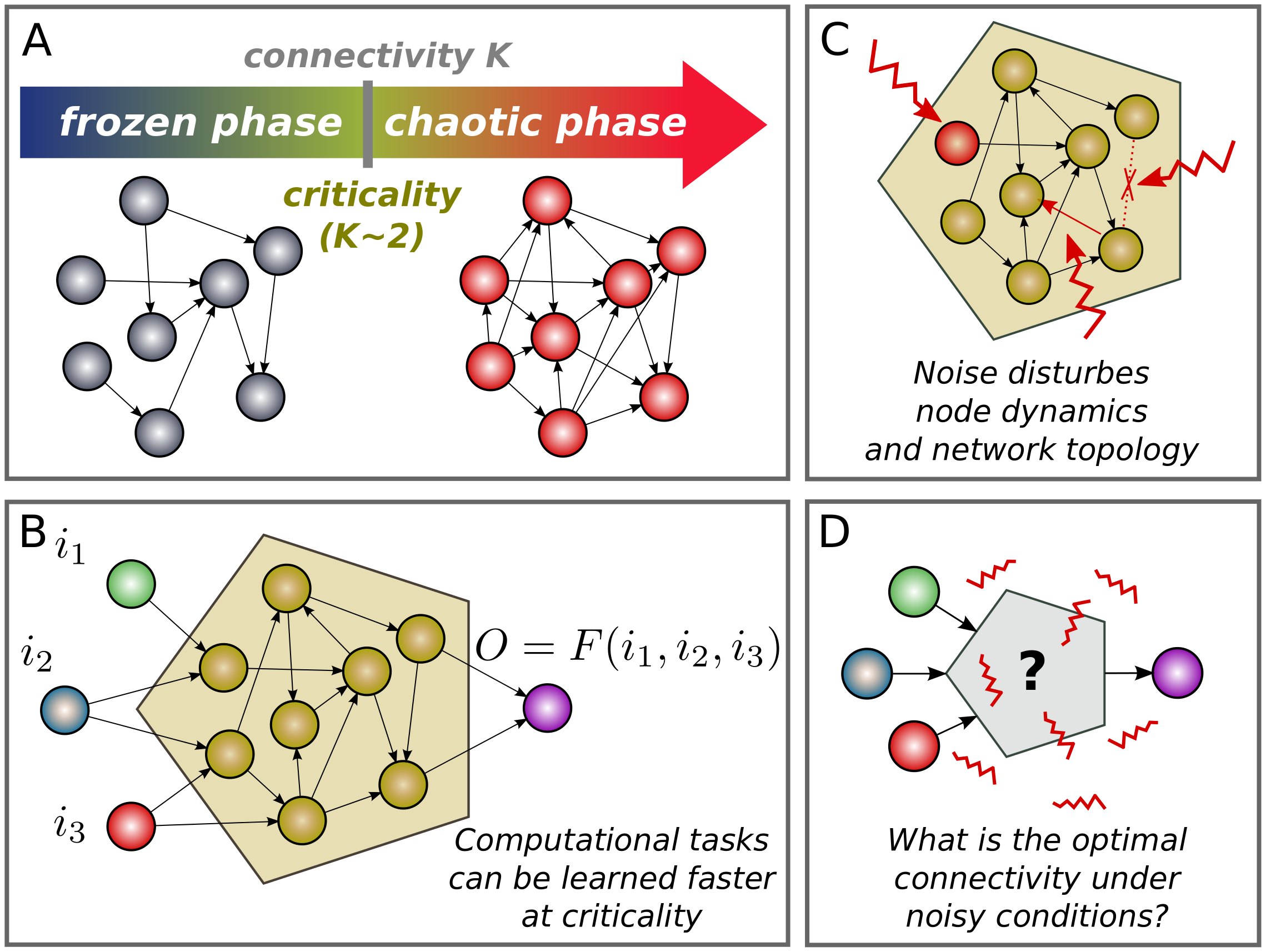} 
 \caption{Sketch of the network architecture. (A) Dynamical phases in general random Boolean networks (RBNs) in the case for which $K_C\approx2$ ($p=1/2$). (B) Constrained network architecture: we impose some ($3$) input nodes (colored in green, blue and red) to receive information from the environment and some output/readout ones ($1$; violet color) to produce a response. 
The overall computational task to be learned can be summarized in a predefined truth table ${\cal{O}}=F(i_1,i_2,i_3)$ where ${\cal{O}}$ is the output state and $i_{1,2,3}$ the input ones.  (C) During the network dynamics and adaptive evolution, there can be noise sources (internal or external) disturbing the network states as well as its topological structure. (D) The aim is to find the optimal connectivity to learn and perform successfully the computational tasks either in the absence of additional stochasticity as well as in the presence of noise.} 
 \end{figure*}

As said above, we consider a setting (similar to that of previous approaches \cite{goudarzi2012}) to train Random Boolean networks to perform a computational task, but we introduce a number of changes --to be detailed in this section-- devoted mostly to implement stochasticity in various ways.

{\bf Network architecture and dynamics.}  Random Boolean networks (RBNs) consist of random Erd\H{o}s-R\'enyi (ER) structures with $N$ nodes, $L$ directed links, and averaged connectivity or degree $K=L/N$.  Self-loops and multiple edges with the same directionality linking two nodes are explicitly excluded, and otherwise the wiring pattern is randomly set. Each node $i$ represents a gene, whose dynamical state is characterized by a Boolean variable $\sigma_i=\{0,1\}$; $1$ for the ``on'' state and $0$ for ``off'' one. The node is updated according to a random Boolean function, $f_i$, which depends on the state of the $K_\mathrm{in}(i)$ neighbor nodes regulating it (restricted to a maximum value of $8$ for computational convenience), and it contributes to regulating the state of $K_\mathrm{out}(i)$ out-neighbors (see Table I in Methods for an example of random Boolean functions). The averaged fraction of $1$'s in the outputs of the random Boolean function, $p$, can be fixed a priori and taken as a control parameter, determining the bias toward ``on'' or ``off'' states (here, we consider the unbiased case $p=1/2$ in all analyses).  In contrast with most studies of RBNs and in order to implement a first source of stochasticity, nodes are updated in an \emph{asynchronous} way \cite{gershenson,drossel2005,peixoto-drossel2010}, i.e. a given node is randomly selected with homogeneous probability, its state is updated according to: \begin{equation}
 \sigma_i(t+\Delta t)=f_i(\sigma_{n^i_1}(t), \sigma_{n^i_2}(t),...,\sigma_{n^i_{K_\mathrm{in}}}(t)),
\label{dynamics}
\end{equation}
where $n^i_j$ identifies the $j-th$ neighbor of node $i$, time is incremented in $\Delta t= 1/N$ units, and the process is iterated. A time step of the dynamics corresponds to one update per node on average.  In order to implement computational tasks or learning rules in RBNs we consider a slight variation of the just-described general architecture, in which some pre-defined input and output nodes are included (see Figure 1B).  By construction, input nodes are imposed to have $K_\mathrm{in}=0$, so that they are not influenced by others and $K_\mathrm{out}>0$, so that they are not isolated, while --on the contrary-- output nodes have $K_\mathrm{out}=0$ and $K_\mathrm{in} \geq 1$ (in particular, we take $n_\mathrm{input}=3$ input nodes and one single output or readout node ($n_\mathrm{output} =1$ as in Fig. 1B).  The set of $N-n_\mathrm{input}$ non-input nodes is called the network core.

{\bf Assessing the network dynamical state.}  In the infinite size limit, synchronous RBNs are known to exhibit a critical point --in the sense of marginal propagation of perturbations \cite{derrida86, gros2011}-- at a value of the connectivity $K_C(p)=\frac{1}{2p(1-p)}$, being ordered/subcritical for $K<K_C(p)$ and disordered/supercritical otherwise.  In particular, in the unbiased case, $p=1/2$, $K_C=2$ (see Figure 1A) which is often quoted as ``the'' critical connectivity for RBNs. However, these results hold only for infinite networks; for finite ones, critical values are shifted toward slightly larger connectivity values by corrections of order ${\cal{O}}(N^{-1})$. Here, instead of calculating such critical values analytically, and thus to quantify possible deviations from criticality, we explicitly compute in numerical simulations the dynamical state of any given finite-size network. For this, 
we determine
whether individual site perturbations do grow or shrink on average; i.e. we measure the branching parameter, $B$, defined as the averaged Hamming distance --after one timestep-- between the original and all possible network-states differing from the original one at just a single (flipped) site (see Methods). Branching parameters $B >1$ (resp. $B<1$) reflect supercritical (resp. subcritical) networks while the marginal case $B=1$ is the trademark of criticality \cite{derrida86, gros2011}.

{\bf Computational tasks.}  The task to be learned can be codified in a ``truth table'', i.e. for each specific input configuration (out of a total of $I=2^{n_\mathrm{input}}$) there is an output value to be reproduced. A given truth table defines a specific computational task. An example is the odd-even classifier (rule R150 in the Wolfram's classification of cellular automata \cite{Wolfram2002}), which assigns a Boolean variable to each input accounting for its parity. Other examples that we consider are rules number R51 and R60 in Wolfram's classification. These rules can be categorized accordingly to their ``complexity'', understanding as such, the number of nodes in the input that do change the output state when altered (and how often they do so for different values of the remaining nodes). In particular, out of the three rules that we study here, the most complex one is the odd-even classifier (R150) whose output obviously depends on all input nodes, R60 is an intermediate case, while the less complex one is R51 whose output is the opposite of one particular input unit, being insensitive to the other two. A more precise definition on how to quantify task complexity --unnecessary for our purposes here-- has been discussed by Goudarzi {\emph et al.} \cite{goudarzi2012}.

{\bf Network fitness.}  The goal of the trained networks is to produce --for each specific input configuration $i$-- a time-averaged value of the output state, $\langle \sigma_\mathrm{output} (i)\rangle$, which is as close as possible to the desired output in the task truth table, $\sigma_\mathrm{output}^*(i)$; the difference between these two values, $\left|\left\langle \sigma_\mathrm{output}\left(i\right)\right\rangle - \sigma_\mathrm{output}^{*}\left(i\right) \right|$, --which is a real number-- is a measure of the network performance for a fixed input configuration. The overall \emph{network fitness} is defined as one minus the average of such difference for $I= 2^{n_\mathrm{input}}$ randomly chosen input configurations:
\begin{equation} F = 1-\frac{1}{I}\overset{I}{\underset{i=1}{\sum}}\left|\left\langle \sigma_\mathrm{output}\left(i\right)\right\rangle - \sigma_\mathrm{output}^{*}\left(i\right) \right|. \label{fitness} \end{equation}

The network is trained to ``learn'' to produce --as fast as possible-- the correct output when exposed to each of the $I=2^{n_\mathrm{input}}$ specific input states; i.e. the network learns the computational task as defined by a given truth table.  To implement this, we sequentially expose the network to $I$ randomly chosen inputs. The resulting random order of inputs can be viewed as a form of stochasticity, mimicking environmental variability. Moreover, the environment is assumed to change rapidly so that, in order to cope with that, networks are trained to reach the correct output within just $t_\mathrm{max}$ (usually fixed to $10$) timesteps, after which the input is changed (while the network state is left unaltered). The first half of this time interval allows for the network to adapt to the new input configuration, while in the second half we measure the average state of the output node $\langle \sigma_\mathrm{output} \rangle$ and compute the value of the network fitness, $F$.

{\bf Network mutations.}  Having established the fitness of a given network, $M$, we now allow it to ``mutate'' by rewiring some existing link --thus preserving its overall connectivity $K$-- and generate a slightly modified network $M'$. The technicalities of how the mutation process is implemented are deferred to the Methods section.

{\bf Network evolution and convergence.}  The network with the largest fitness value, between $M$ and its mutated counterpart $M'$, is selected (while the original one is kept if the two fitnesses coincide).  This mutation and selection process defines an \emph{evolutionary time step} (to be distinguished from a time step of the dynamics; there is a factor $t_\mathrm{max} I$ between both). The evolutionary process is iterated until $F$ reaches its maximal possible value $F=1$. Observe, however, that as the $I$ inputs are randomly chosen at each evolutionary step, observation of $F=1$ at a given step does not necessarily imply $F=1$ at successive time. Therefore, in order to impose that the network robustly ``learns'' the computational task, we continue to measure its fitness, when exposing it to a much large number of randomly chosen inputs ($100 I$, instead of just $I$ as in the fitness-computation Eq.(\ref{fitness})); if $F=1$ all accros this long checking time window, the network is classified as having learned. Otherwise, the mutation/selection process is restarted until an optimally performing network is found. The final number of evolutionary steps required to reach an optimal network is called \emph{convergence time}, $T$.

{\bf Ensemble averages.}  Keeping fixed specific values of the network size $N$ and connectivity $K$, the previous evolutionary process is iterated a large number of times (typically from $10^3$ to $5\cdot10^5$) giving rise to an ensemble of trained networks.  The ensemble averaged convergence time, $\bar{T}=\bar{T}(N,K)$, is a proxy for the network performance: the best network ensemble is the one with the smallest $\bar{T}$. In this set of networks --once they have been trained-- we also measured the ensemble-average of the branching parameter, $\bar{B}$.  In the approach of Goudarzi \emph{et al.} \cite{goudarzi2012}, $K$ is allowed to change during the evolutionary process; thus the fastest learning networks are selected for; instead, we explore different fixed-$K$ ensembles and determine a posteriori which is the optimal one.  Both approaches are obviously equivalent to determine the optimal connectivity $K$.

{\bf Dynamics under noisy conditions.}  To investigate the effect of fluctuations in the system dynamics, we allow the dynamics to be exposed to noise. In particular, we consider that either (i) with a small probability, $\eta$, nodes can invert their state every time they are updated (accounting for errors/fluctuations in gene expression levels) or (ii) with some small probability, $\xi$, (which is proportional to the network connectivity) the network topology experiences a mutation process at each evolutionary step, and the mutated network is kept/selected regardless of its fitness value (this describes physical damage in the network produced, for example, by the lack or excess of some regulatory factors).  For the sake of simplicity, we refer to the first possibility as ``dynamical'' noise and to the second one as ``structural'' noise.

\section*{Results}
\subsection*{Convergence times and dynamical phases of learning networks}

Even in the absence of explicit noise sources, the dynamics based on asynchronous updating --which is the one we adopt here-- has a stochastic component (i.e. nodes are updated in a random order), which could be more adequate to represent real genetic networks than synchronously updated RBNs as it avoids spurious effects associated with perfectly synchronous updating \cite{drossel2005}.

We consider a complex computational task --the odd-even classifier-- and analyze networks of variable $N$ and $K$. We let them evolve to learn this task and measure the average convergence time, $\bar{T}$, to do so.  Results are shown in Figure 2 for sizes from $N=6$ to $N=64$ as a function of the network connectivity $K$ (from $K=0.5$ to $K=3.5$). First of all (upper Fig. 2B), observe that for all values of $N$, $\bar{T}$ exhibits a characteristic (pseudo)parabolic shape with a minimum at some optimal connectivity value, $K_T$, at which networks learn the computational task in the fastest possible way. It is important to stress that networks with connectivities other than $K_T$ also learn, even if after longer evolutionary times. In Fig. 2A the same data are represented, but rescaling $\bar{T}$ for each $N$ with its minimum, $\bar{T}_\mathrm{min}(N)$ (this is done to help the eye to compare the location of the different minima). In Fig. 2C we plot $|K_T-2|$ as a function of $N$ (blue squares); the value $K=2$ corresponds to the usually accepted critical connectivity for RBNs in the infinite size limit. Observe that the optimal connectivities seem to converge to this value, $K=2$, as a power-law function of $N$.  The precision of our numerics does not allow us to discriminate if the convergence is exactly to $K=2$ or to a nearby value (within $2.00\pm 0.05$) in the large-size limit. In Fig. 2A, we also present results for the branching parameter, $\bar{B}$ (see Methods), for the same network ensembles, which allows us to explicitly determine
the average dynamical regime as a function of $K$.  Importantly, $\bar{B}$ is computed in the ensemble of networks that have learned --and not in the Erd\H{o}s-R\'enyi ensemble-- and Hamming distance measurements are restricted to the network core (excluding input nodes, which do not change in the course of the dynamics).  In particular, dotted lines in Fig. 2A stand for measurements of $\bar{B}$, after perturbing nodes in the core, while dashed-dotted lines correspond to perturbations at input nodes.  Observe that these two sets of curves exhibit qualitatively different behaviors.  We have chosen to present results in this way to stress the fact that --after learning-- networks are not homogeneous, and not all nodes respond in the same way; in particular, the network is more responsive (larger $\bar{B}$) to input perturbations than to changes in the core.  For example, networks with connectivity $K=2$ are supercritical to input perturbations (fostering network sensitivity to external changes) and subcritical for core perturbations (as required for a robust convergence to the attractor/output).

\begin{figure}[H]
\centering \includegraphics[width=12cm,angle=0]{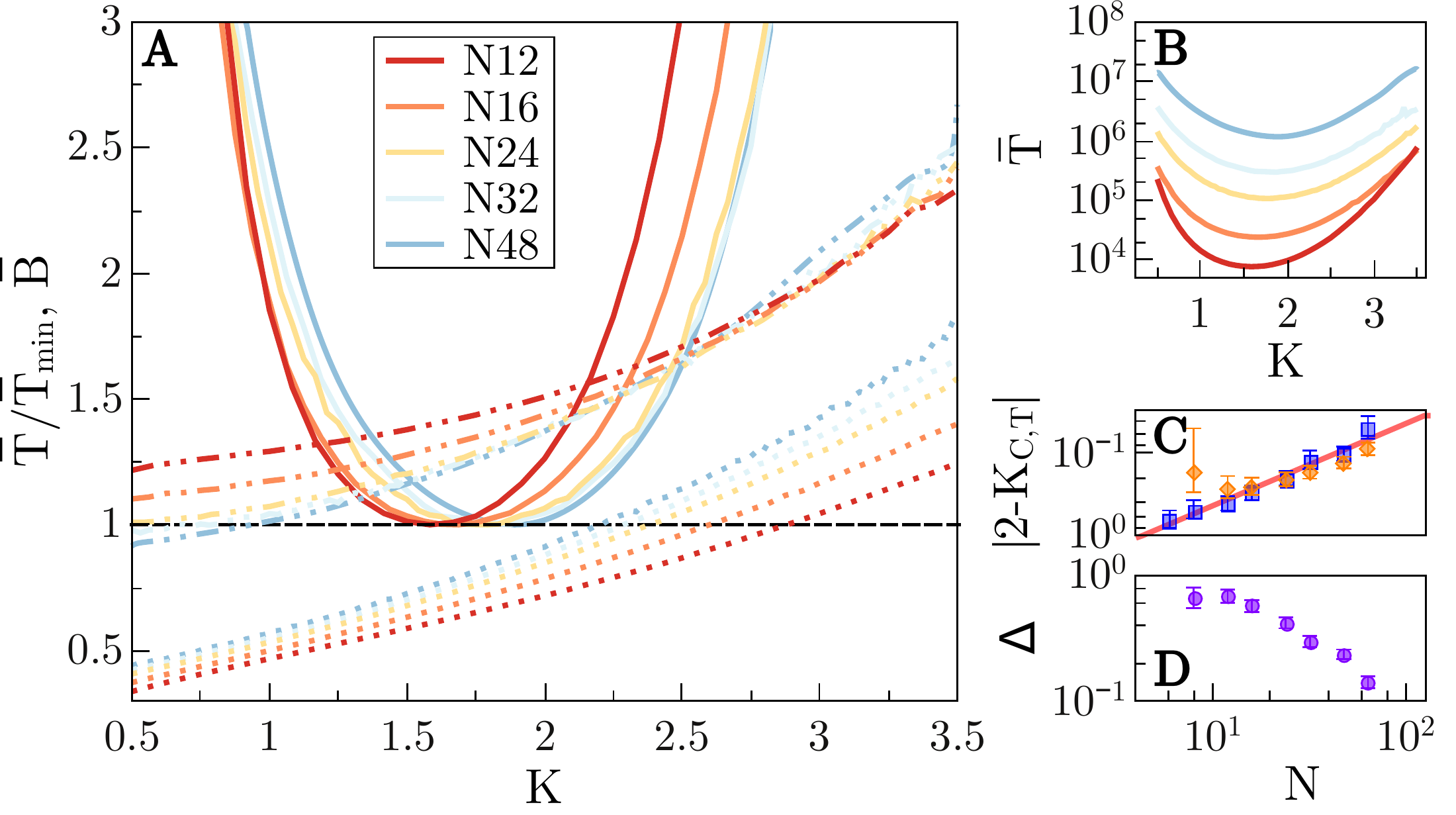}
\caption{(A) Averaged convergence time, $\bar{T}$, divided by its minimum value $\bar{T}_\mathrm{min}$ plotted as a function of the network connectivity $K$ for different network sizes $N$ (solid lines); ensemble averages performed over up to $5\cdot10^5$ network realizations. Similarly, the non-normalized convergence times are plotted in (B).  Discontinuous lines in (A) represent the value of the branching parameter, $\bar{B}$ as measured in the network after the learning process is completed; dashed-dotted lines stand for $\bar{B}$ averaged after perturbing only input nodes, while dotted lines have been obtained after perturbing nodes in the network core. Note that as $\bar{T}/\bar{T}_\mathrm{min}$ and $\bar{B}$ are both dimensionless quantities, they have been plotted in the same scale; the same color code has been used for all curves. (C) Scaling of the connectivity at which the minimum $\bar{T}$ is obtained, $K_T$, as a function of $N$ (blue squares), plotted together with the position of the critical point $K_C$ as estimated from the condition $\bar{B} \approx 1$ (orange diamonds). In both cases, there is a convergence toward a value close to $2$ in the large $N$ limit (blue squares) (the red line is a guide to the eye and corresponds to a decay ${\cal{O}}(N^{-1})$ toward $2$. (D) $\Delta = K_C -K_T $ plotted as a function of $N$ showing explicitly that the distance to criticality diminishes with network size; i.e. the larger the network the closer to criticality the fastest learning networks. }
 \end{figure}

To obtain the overall branching parameter $\bar{B}$ (given $N$ and $K$) --for all nodes in the network-- we need to average these two contributions (weighted with $n_\mathrm{input}=3$ and $N-3$ nodes, respectively). For these averaged curves (which, for the sake of clarity, are not explicitly shown in Fig. 2A) the crossing $\bar{B}=1$ indicates overall critical dynamics, and corresponds to a critical connectivity $K_C$. $K_C$ turns out to be larger than $K=2$ and shifts toward lower connectivity values as $N$ grows; indeed, its distance to $K=2$ decreases with $N$ (see Fig. 2C; orange diamonds), suggesting that learning networks have critical connectivity $K\approx 2$ (within our resolution) in the infinite size limit, as happens with random networks.

Moreover, we have measured the difference $\Delta = K_C -K_T $ to gauge how far optimal connectivities (in the sense of achieving the fastest possible learning) are from critical dynamics (in the sense of the branching parameter as close as possible to $1$). As shown in Fig. 2D (magenta circles), $\Delta$ decreases monotonically upon increasing $N$, indicating that --for sufficiently large networks-- the optimal connectivity is as close to criticality as desired, but for any finite size they are slightly subcritical ($\Delta >0$). Thus optimal learning occurs for slightly subcritical networks,  arbitrarily close to criticality for sufficiently large system sizes. 

Figure 3A illustrates results for other, less complex (see above) computational tasks. As before, there is a well-defined minimum for $\bar{T}$ in all cases, but these times are significantly shorter for lesser complex tasks (about two orders of magnitude less for a fixed size). Observe also that for the simplest, $R51$ rule, $\bar{T}$ hardly depends on $K$ (Fig. 3C), indicating that, as the task complexity decreases $K$ progressively becomes a lesser relevant parameter.  Observe also (Fig. 3D) that the distance of optimal networks to criticality, $\Delta$, decreases with increasing network complexity. Therefore, it is reasonable to conjecture that for more complex tasks than the ones we considered (e.g. involving larger values of $n_\mathrm{input}$), the benefits derived from operating at optimality/criticality are progressively more crucial.

\begin{figure}[H]
\centering \includegraphics[width=12cm,angle=0]{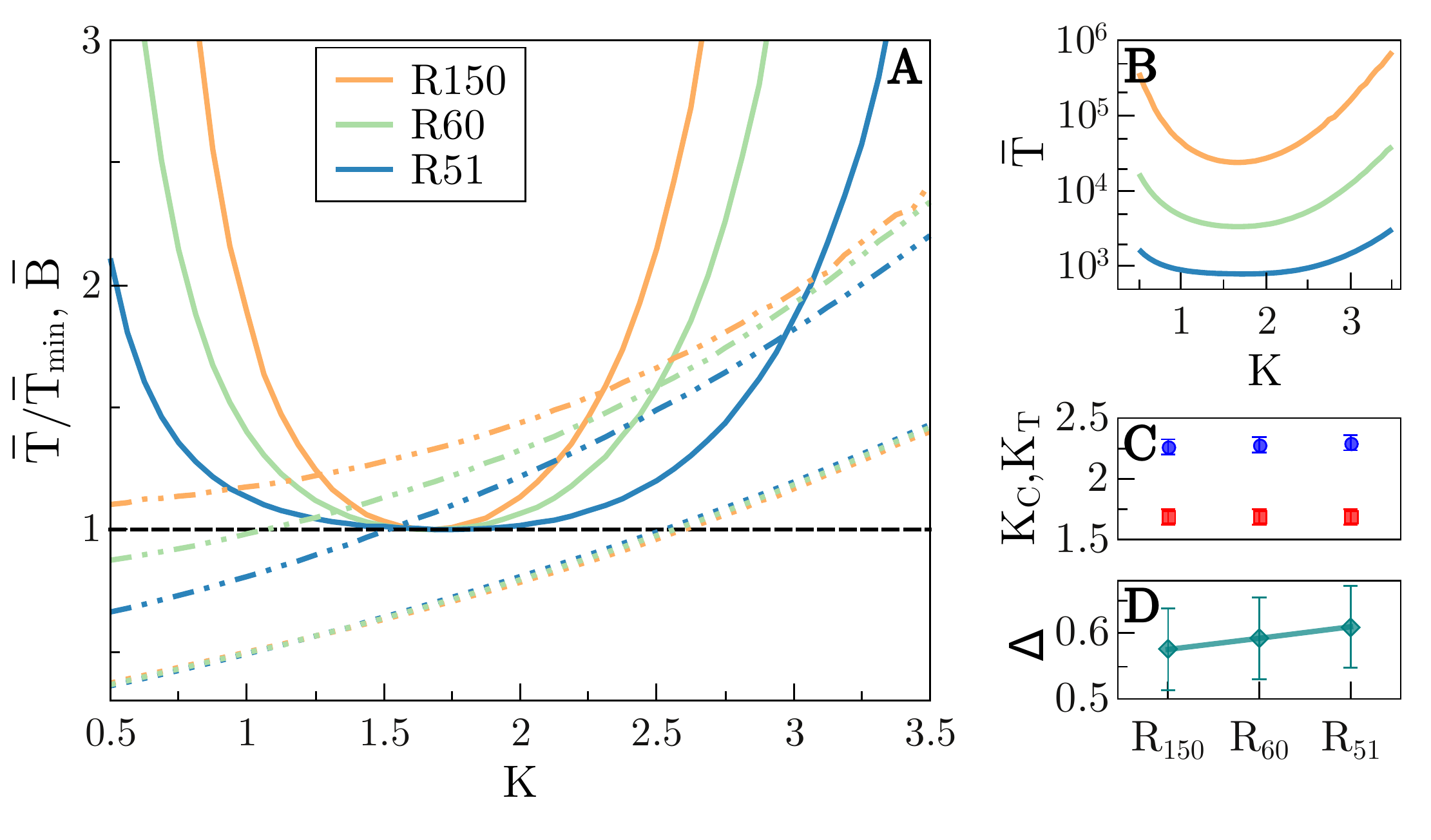}
\caption{Figure analogous to Fig. 2, but obtained for different computational tasks of different complexity (from the most complex R150 to the intermediate R60 and the simplest R51 rule; the names come from Wolfram's classification of cellular automata \cite{Wolfram2002}).  (A) $\bar{T}/\bar{T}_\mathrm{min}$ for $N=16$ (solid lines) and the three considered rules. Discontinuous lines are as in Fig. 2. (different colors stand for different rules). (B) Non-normalized $\bar{T}$ for $N=16$; the same color code has been used for all curves. (C) Optimal-time connectivities for fast learning, $K_T$ (red squares) and critical connectivities $K_C$ (blue circles) for the different rules. Observe that in all cases, optimal networks are slightly subcritical for this relatively small size $N=16$.  As shown in (D) the distance to criticality decreases upon increasing the task complexity.}
 \end{figure}

Finally, we also scrutinized the network topology (in-degree distribution) after learning and,
 interestingly, we did not detect significant structural changes, as the overall network skeleton was in all cases  very close to a random network.

Summing up, in order to achieve the fastest possible learning of complex tasks, RBNs with a connectivity such that their dynamics turns out to be critical (or slightly subcritical for finite sizes) are the best possible option. \emph{The larger the network size and the more complex the task, the more evolutionarily favourable to be close to criticality}. 

\pagebreak

\subsection*{Learning under noisy conditions}

\subsubsection*{Dynamical noise}
Figure 4 is analogous to Fig. 2 but has been obtained in the presence of dynamical noise, $\eta \neq 0$ (results for $\eta =0$ are also plotted for the sake of comparison); observe that we present results for a fixed size $N=16$ and variable noise strengths (from $\eta=10^{-5}$ to $\eta=10^{-3}$). It is noteworthy that for larger values of $\eta$ (e.g. $0.01$) the dynamics is so noisy that the probability for the networks --resulting out of the evolutionary process-- to pass the robustness filter we have imposed (i.e. to have fitness $F=1$ for $100 I$ evolutionary steps) is exceedingly small. Therefore, networks do not achieve perfect learning in such extremely noise conditions.  On the other hand, for exceedingly small noise strengths, we essentially see the same results as for $\eta=0$, within the simulation checking time windows we consider.  For intermediate noise-strength levels (such as the ones reported in Fig. 4) networks are likely to pass the filter. In such cases, (see Fig. 4B), the optimal connectivity is observed to shift toward lower values of $K$ as the noise level is increased (see also Fig. 4C where $K_T$ is plotted as a function of $\eta$ for various system sizes). In parallel, the averaged convergence times, $\bar{T}$ (Fig. 4B, same color code as in panel A), also grow with noise.

\begin{figure}[H]
\centering \includegraphics[width=12cm,angle=0]{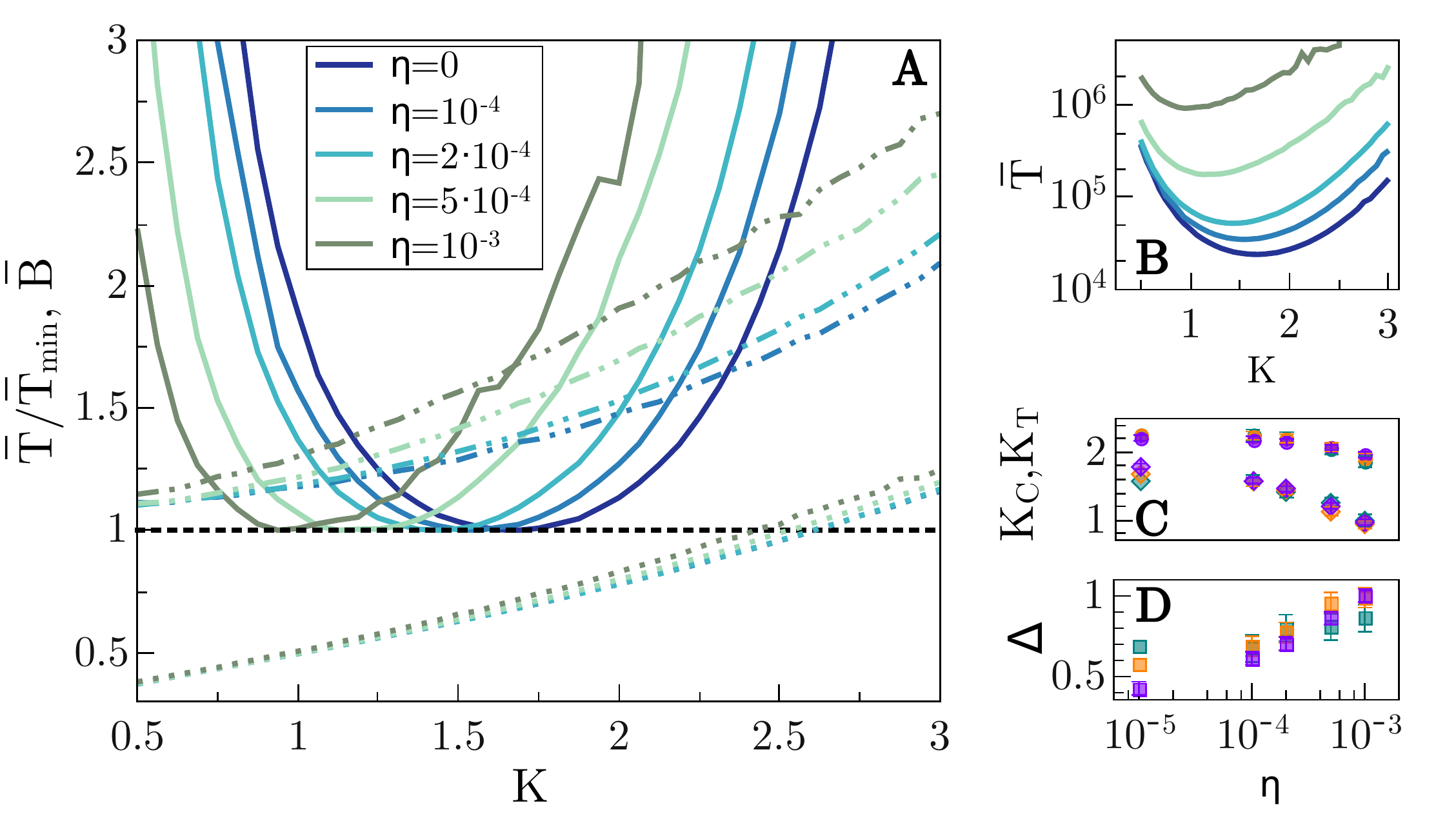}
\caption{Figure analogous to Fig. 2. and Fig. 3 but for analyzing the dependence on the strength $\eta$ of the dynamical noise. (A) $\bar{T}/\bar{T}_\mathrm{min}$ for $N=16$ (solid lines) as a function of $K$ for different values of $\eta$ (different colors). Discontinuous lines are as in Figs. 2 and 3 (however, different colors stand now for different $\eta$ values). (B) Non-normalized $\bar{T}$ for $N=16$; the same color-code has been used for all curves. (C) Optimal-time connectivities for fast learning, $K_T$, (diamonds) and critical connectivities $K_C$ (circles) for the different values of $\eta$ and various network sizes ($N=12$ blue, $N=16$ orange, $N=24$ purple).  In all cases, optimal networks are slightly subcritical for this relatively small sizes. However, in contrast with the noiseless cases above, here (D) the distance to criticality $\Delta$ does not decrease upon enlarging the size (except for exceedingly small noise strengths, e.g. $10^{-5}$, for which noise effects are not visible in the time windows we consider) actually it remains almost constant or --for large values of $\eta$ such as $10^{-3}$-- it grows with $N$, and in any case, it grows with the noise strength (same color code used in C and D).}
 \end{figure}

On the other hand, the branching parameter (measured keeping  the noise  switched on) computed by perturbing core nodes does not show a strong dependence on $\eta$ (see dotted lines in Fig. 4A) while the values of $\bar{B}$ obtained by perturbing just the inputs (dashed-dotted lines in Fig. 4A) are more severely affected.  The resulting critical points obtained by averaging these two contributions and equating them to unity are plotted in Fig. 4C, are always close to $K=2$ (for the considered sizes). Comparing these values with the optimal connectivities for learning, i.e. measuring, $\Delta = K_C -K_T $, one observes (see Fig. 4D) that $\Delta$ increases monotonically with $\eta$. This occurs for the different system sizes we studied allowing us to conclude that \emph{ under noise conditions, it takes longer to learn, and the larger the dynamical-noise strength the more subcritical the optimal networks}.

\subsubsection*{Structural noise}
Figure 5 shows results analogous to those in Fig. 4. Also in this case we present results for a fixed size $N=16$ and variable noise strengths (from $\xi=10^{-3}$ to $\xi=10^{-2}$). In parallel with the site-noise case, there is a noise intensity threshold above which the mutation probability is exceedingly high for the networks to learn, while for too small strengths, the same results as for $\xi=0$ are observed within the operational checking time windows we have.  For intermediate noise amplitudes, the larger $\xi$ the longer the learning process takes (see Fig. 5B). In these cases, the optimal connectivity is observed to shift toward lower values of $K$ as the noise level is increased (see also Fig. 5C where $K_T$ is plot as a function of $\xi$). Also, as above, the branching parameter, $\bar{B}$ (measured keeping a fixed network structure) does not have a strong dependence on $\xi$ (Fig. 5A). The associated critical point $K_C$ is slightly above $K=2$ for small noises, and moves progressively to smaller connectivity values as $\xi$ grows.  Also, as in the previous case, $\Delta$ increases monotonically with $\eta$, so that, as above, \emph{we can safely conclude that, in general, the larger the structural noise strength the more subcritical the optimal networks}.

\begin{figure}[tb]
\centering \includegraphics[width=12cm,angle=0]{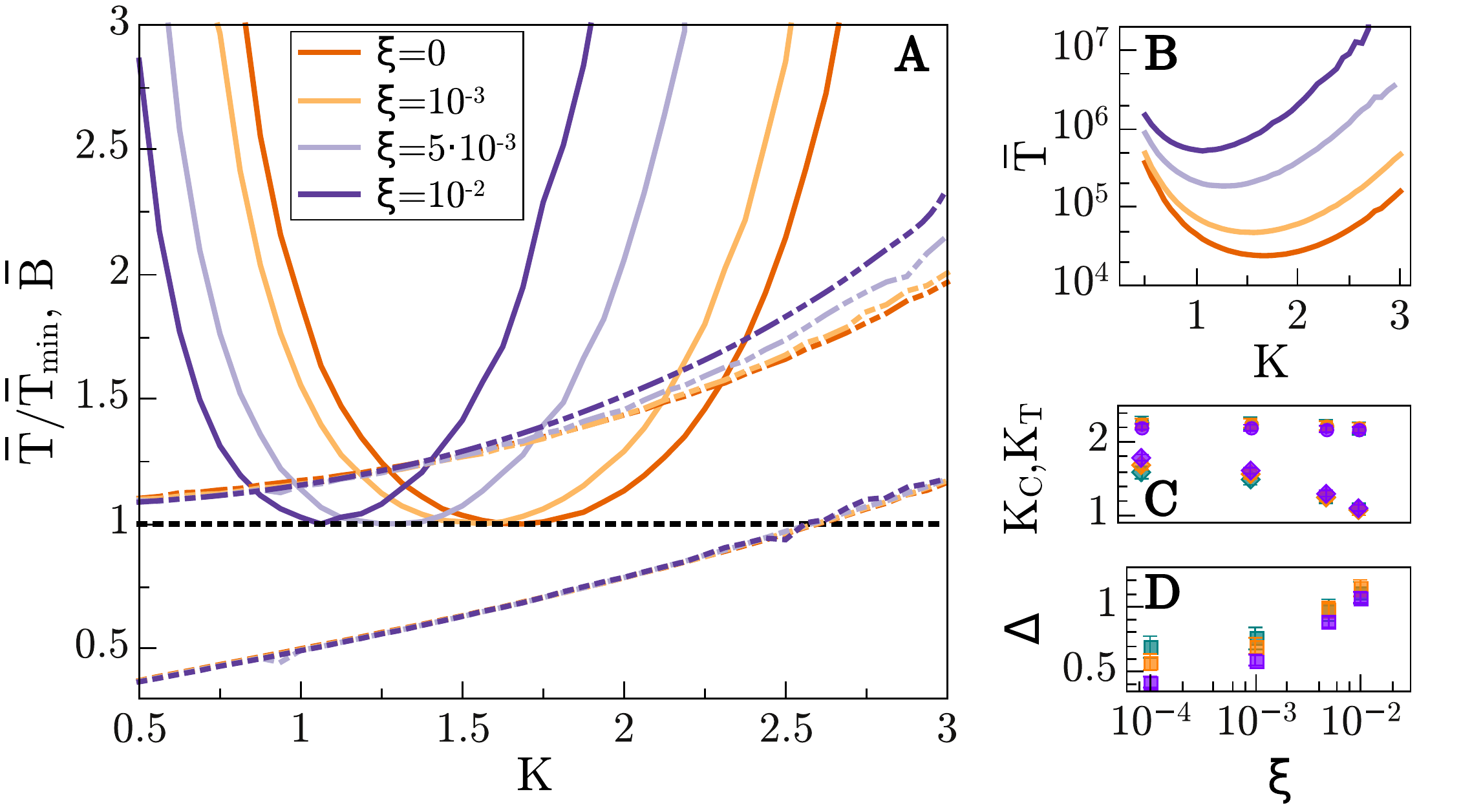}
\caption{Figure analogous to Fig. 4. but analyzing the dependence on the strength $\xi$ of the structural noise. (A) $\bar{T}/\bar{T}_\mathrm{min}$ for $N=16$ (solid lines) as a function of $K$ for different values of $\xi$ (different colors). Discontinuous lines are as in Fig. 4. (different colors stand now for different $\xi$ values). (B) Non-normalized $\bar{T}$ for $N=16$ (C) Optimal-time connectivities for fast learning, $K_T$ (diamonds) and critical connectivities $K_C$ (circles) for the different values of $\xi$ and various network sizes ($N=12$ blue, $N=16$ orange, $N=24$ purple).  Observe that in all cases, optimal networks are slightly subcritical for this relatively small sizes. However, in contrast with the noiseless cases above, and in parallel with the case of dynamical noise, here (D) the distance to criticality $\Delta$ does not decrease upon enlarging the size (except for extremely low values of the noise, as in Fig.4), actually it remains almost constant and, in any case, it grows with the noise strength. Same color code used in C and D.}
 \end{figure}

Summing up, we conclude that while in the case of noiseless dynamics the optimal solution --to achieve the fastest possible learning-- is obtained at connectivities for which the network is about critical (actually slightly subcritical, but closer and closer to criticality as the network size and/or the complexity of the task are increased), the situation is different in the presence of additional stochasticity, be it dynamical or structural noise. Under noisy conditions, the optimal solutions lie clearly well within the ordered/subcritical phase. A straightforward interpretation of this result is that the network dynamics needs to compensate for the excess of noise, and does so by reducing its internal level of uncertainty, i.e. by shifting deep into the ordered/subcritical phase.

\subsection*{Empirical networks}
We have collected a set of empirical data from the literature and compiled a set of real directed networks. This includes public empirical datasets with biological genetic regulatory networks \cite{stark2006}, and networks of metabolic interactions \cite{feist2009}.  Specific examples of networks collected from the literature are the metabolic networks of \textit{Chlamydomonas reinhardtii} ($K=2.05$) \cite{chang2011}), and \textit{Bacillus subtilis} ($K=1.03$) \cite{oh2007}, and the gene regulatory networks of \textit{Escherichia coli} ($K=1.24$, $K=2.32$) \cite{balazsi2005, peixoto2012}, \textit{Arabidopsis thaliana} ($K=2.755$) \cite{ma2007}, \textit{Mycobacterium tuberculosis} ($K=1.19$, $K=1.98$) \cite{balazsi2008, sanz2011trans}, \textit{Pseudomonas aeruginosa} ($K=1.48$) \cite{galan2011}, and \textit{Saccharomyces cerevisiae} ($K=1.85$) \cite{guelzim2002}. Figure 6 presents a scatter plot of all networks in our dataset, representing the averaged connectivity $K$ and network size $N$ of each one. As it can be seen, the averaged connectivity of this dataset is well below the value $K=2$, the critical connectivity for large random networks, suggesting that they could operate in subcritical regimes. It is noteworthy that it has been suggested that some empirical networks with high connectivity values (such as some of the outliers in Fig. 6) might result from systematic errors in correlation analyses (giving rise to false positives) \cite{leclerc2008}.

Being more precise --given the absence of knowledge on dynamical aspects of the specific dynamics of each empirical network-- it is not possible to properly ascertain the dynamical state (critical or not) of each of them.  For instance, in large random Boolean networks the critical point is located as discussed above at $K_C=\frac{1}{2 p(1-p)}$ \cite{gros2011,drossel2008,aldana-review}; thus the minimal possible critical connectivity is $K=2$ (corresponding to the unbiased case $p=1/2$.  Note that for finite random networks, the critical connectivity shifts to values slightly larger than $2$ (positive corrections of order $N^{-1}$).  Therefore, if the collected (finite) empirical networks obeyed random Boolean dynamics-- in light of Figure 6-- almost all of them would be certainly subcritical. However, we know that the dynamics of real networks may involve, for instance, canalizing and/or weighted updating functions \cite{drossel2008}, and for such networks the critical connectivity can be in some cases smaller than $K=2$. Therefore, even if no definite conclusion can be extracted from these empirical data about the possibility of criticality (or absence of it), we can certainly conclude that empirical networks are quite sparse (significantly sparser than critical random networks) suggesting that --in the absence of further information about their intrinsic dynamics-- the most likely scenario would be that they operate in ordered regimes (see below for an extended discusion).

\begin{figure}[H]
\centering \includegraphics[width=9cm,angle=0]{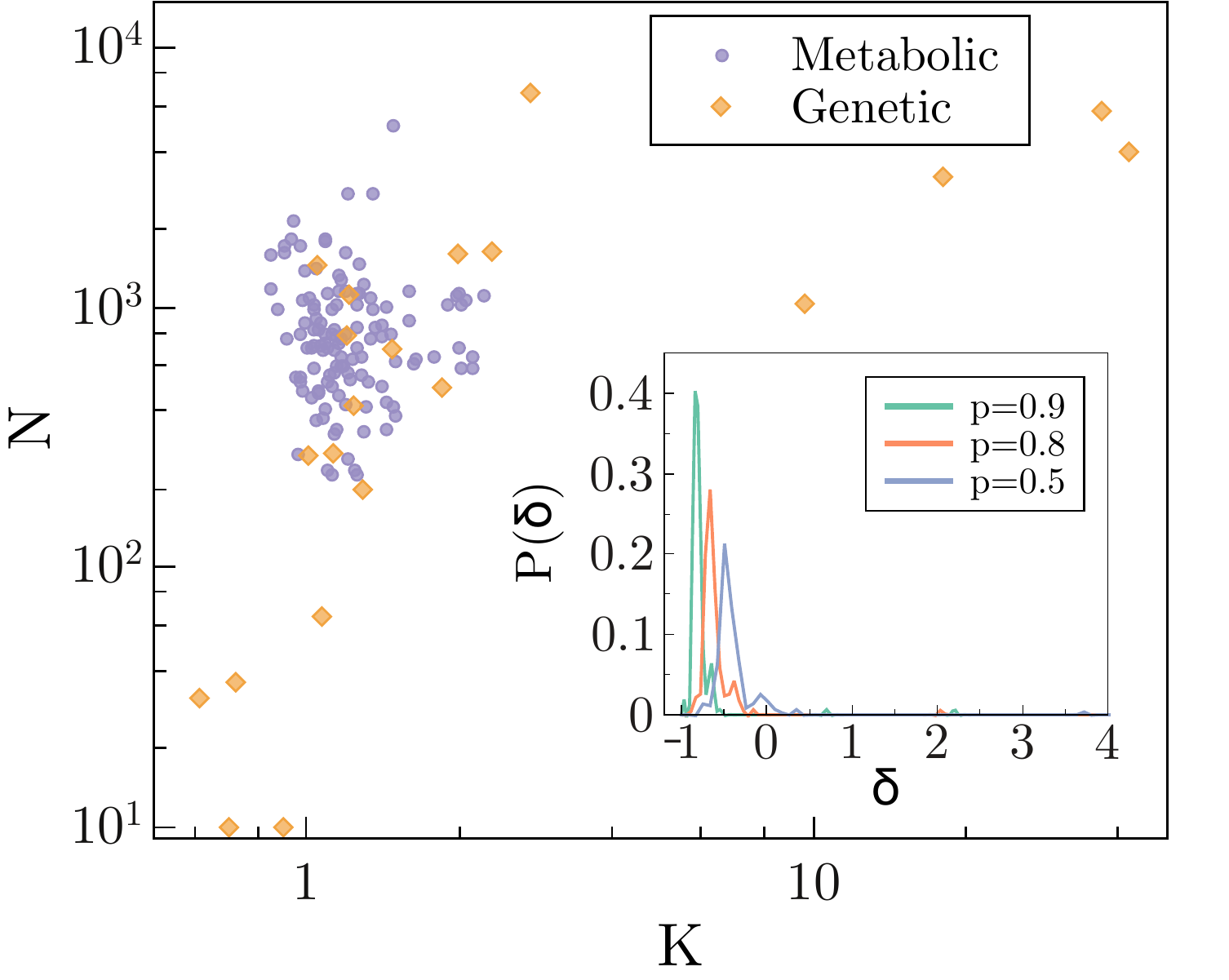}
\caption{Size $N$ versus mean connectivity $K$ for empirical biological networks of different types.  In particular, it includes genetic and metabolic networks of species such as \textit{Escherichia coli}, \textit{Saccharomyces cerevisiae}, \textit{Pseudomonas aeruginosa}, \textit{Bacillus subtilis}, etc (see main text).  Observe that all networks are significantly sparse, with most mean connectivities lying between $K=1$ and $K=2$. The outliers, with $K>10$ come all from BioGRID \cite{stark2006}; the most extreme case has $K= 41.90$ and corresponds to the genetic network of \textit{ ``Escherichia coli K-12 W3110''} (but, it might be that these networks are plagued with false-positive connections \cite{leclerc2008}).  In the inset, we plot the probability that a network from our empirical ensemble is at a certain relative distance to the critical point of a random Boolean model with its corresponding connectivity, i.e. $\delta=(K-K_c(p))/K_c(p)$,
 assuming a fixed value of the bias $p$ (in particular, we show results for $p=1/2$, $p=0.8$ and $0.9$); observe that regardless of the value of the considered bias (which in general is unknown to us) most of the networks lie within the subcritical regime (assuming their dynamics was random).}
 \end{figure}

\section*{Conclusions and Discussion}  The hypothesis that living systems may operate in the vicinity of critical points of their internal dynamics has inspired and tantalized scientists for some time. In particular, it has been claimed that genetic regulatory networks might operate close to criticality, achieving in this way an optimal balance between sensitivity to signals and stability to noise, and/or between adaptability and robustness on large evolutionary scales.  A few works have recently explored different mechanisms allowing for networks to self-organize or evolve to critical or quasi-critical dynamics.

Here --inspired by the set up proposed by Goudarzi et al. \cite{goudarzi2012}-- we have shown that random Boolean network models that are trained to perform a given computational task, can learn it much faster if they have a connectivity $K$ such that their dynamics turns out to be close to criticality, as defined by a marginal averaged propagation of perturbations. This does not mean that networks far from criticality cannot learn; indeed they do, but it takes much longer to do so. Two important differences between the present work and previous ones are as follows.  First, we work with networks with constant connectivity, i.e. the allowed mutations keep $K$ constant, while in previous work there was no such constraint \cite{goudarzi2012}.  This difference implies that our evolutionary process does not converge to the optimal connectivity for fast learning, $K_T$; by studying the constant-connectivity ensemble, we are able to put forward that learning (not necessarily in the fastest possible way) is compatible with rather diverse connectivity patterns and, thus, with the network being critical, subcritical or supercritical.  The second important difference is that we implement a stochastic updating scheme, which introduces stochasticity in the dynamics; we find, however, that results are mostly insensitive to this change. Moreover, we have seen that in all cases, the distance to criticality of the optimal-connectivity networks diminishes monotonically upon enlarging system size and upon enlarging the task complexity. Indeed, very simple tasks, establishing simple relationships between (a few) inputs and the output, can be readily learned by networks in the ordered/subcritical regime, where such a direct correspondence can be robustly realized. On the other hand, complex tasks, in which the output is sensitive to many different possible changes in the input nodes, require of much larger responsiveness/susceptibility, and thus, shift the network optimal connectivity toward larger values, closer and closer to criticality. In any case, we do not find under any circumstances the optimal connectivity to lie within the disordered/supercritical regime; it seems as if the requirement to learn a task was incompatible with the network being disordered.

Biological systems must have homeostasis, i.e. the capacity to maintain their internal conditions even in the presence of fluctuations and noise, be it internal or external.  In the second part of our study we posed ourself the question of how do these results depend upon the explicit introduction of noise.  To this end, we have introduced more extreme forms of noise, be it dynamical or structural, within the same RBN model. 
Dynamical noise allows network nodes to invert their dynamical state with a small probability each time they are updated, introducing perturbations that can potentially propagate through the system, compromising the network performance. Similarly, structural noise, implying that the network topology itself is exposed to random changes with some small probability, also producing potential damage in the learned patterns.  Both of these noise sources have clear correspondence with stochastic effects in real biological networks. In both cases, there is a threshold in noise strength above which networks do not learn the computational task in a reliable and robust way; i.e. they end up being plagued with errors, hindering network learning.  Such thresholds clearly depend on the criterion imposed to declare that networks have learned; put differently, if the time in which one checks for network robustness are increased, i.e. if the criterion becomes more stringent, the noise-strength thresholds diminish.  Remarkably, in both of the cases, dynamical and structural noise, we find that the optimal connectivity to achieve the fastest possible learning lies deep-inside the subcritical region, far away from criticality, and the distance to criticality increases upon enlarging the noise strength and does not diminish upon increasing the system size (as it happens in the absence of explicit noise).

Our results suggest that real biological networks, in order to perform the complex tasks required for information processing and survival in a noisy world, should operate in sub-critical regimes rather than in critical ones as it has been argued. As a matter of fact, the collection of empirical (genetic and metabolic) networks that we have compiled from the recent literature shows a rather sparse averaged connectivity in most cases, with only a few outlier networks. If the dynamics underlying these networks could be modeled by random Boolean functions, one could safely conclude that they are typically subcritical. However, in most cases, the dynamics remains mostly unknown, and a clear cut conclusion about the dynamical state of each specific network instance cannot be derived. To fill this gap, recent analyses have employed high throughput data from hundreds of microarray experiments to infer regulatory interactions among genes. This type of approach leads to more detailed information on dynamical aspects (e.g. switching off a given gene it is possible to follow the cascade of modifications it generates through the whole network).  The resulting data, implemented into Boolean models, seem to support the hypothesis that regulatory networks for a number of species (\textit{Saccharomyces cerevisiae}, \textit{Escherichia coli}, etc) are close to criticality \cite{balleza2008, nykter2008}, but some other analyses leave the door open for the networks to operate in an ordered/subcritical phase \cite{shmulevich2005,kauffman2003}.  Therefore, given the present state of affairs, one can only conclude that more accurate and extensive experimental approaches (including, in particular, more accurate direct measurements of the bias $p$) would be extremely valuable to shed further light on this fascinating problem.

An important observation to be made is that the tasks we have employed to be learned are relatively simple (as they only involve a maximum of $3$ input nodes and a single readout). Thus, one can wonders what would happen if a more extensive use of the network potentiality was necessary (by employing for instance, two or more tasks simultaneously, and/or involving a much larger number of inputs in each single task). Under the light of our results for the noiseless case --where we found that upon considering far more complex tasks, involving many more input and output nodes, the dynamics becomes progressively more critical-- it would not be surprising that if one could analyze much more complex tasks --as the ones probably controlling real biological networks--
 the dynamics could become closer to criticality even in the presence of noise.  Furthermore, in such more complex cases, one should also relax the criterion to declare that networks have learned, and look for ``fuzzy'' types of learning (i.e. accept networks with fitnesses slightly smaller than one). The combination of much more complex rules together with less rigid criteria for learning, could very likely shift the optimal solutions toward more critical states. A detailed analysis of these issues is left as an open challenge for future work.

It is also noteworthy that --even if network topology is known to play a very important role in the outcome of RBNs \cite{aldana2007, Aldana2012,drossel2009, drossel2010, aldana2003, leclerc2008}, here we have focused mostly on random Erd\H{o}s-R\'enyi networks and left the analysis of important topological features of empirical networks --such as scale-free connectivity distributions, and hierarchical and modular organization-- for future work. These aspects might also play an important role in determining the network dynamical state.  Finally, we also plan to extend the studies beyond the limit of the Boolean approach and to implement more complex and biologically realistic tasks.  Hence, our summary is that the criticality hypothesis remains as a valid and fascinating possibility, but that it needs to be critically evaluated under each set of specific circumstances, avoiding making exceedingly general claims.

\section*{METHODS}
 \subsection*{Network mutations.}
\begin{enumerate}
\item Given a original network, $M$, we perform a rewiring, which consists in choosing a link (say from node $i$ to node $j$), removing it, and introducing a new one (from $i$ to $j'$) assuming this one did not exist before (and keeping the topological constraints described above).

\item This change of the network topology, requires some modifications in the random Boolean functions $f_j$ and $f_{j'}$ (see Table 1).  For $f_j$ one needs to eliminate the input $\sigma_i$; thus $f_j$ changes from being a function of $K_\mathrm{in}(j)$ arguments to a function of $K_\mathrm{in}(j)-1$. The new function coincides with the original one fixing $\sigma_i=0$, i.e. for the case when the driving node $i$ was off.  After this, each output in its table is changed with probability $1/4$, defining the ``mutated'' Boolean function. Similarly, for node $j'$ a new argument, $\sigma_i$, is introduced to the Boolean function $f_{j'}$: all values for $\sigma_i=1$ (``on'' $i$ node) are assigned randomly, while for $\sigma_i=0$ (the new input is off) we keep the pre-existing Boolean-function values.

\item This whole rewiring process is performed the first time with prob. one; after that a second rewiring is attempted with prob. $1/2$; if it occurs, then a third one happens with prob. $1/3$ and so on, giving rise to a mutated network, $M'$. This sequential process allows for the possibility of large mutations, involving many re-wirings.

\item Observe that these mutations keep the out degree sequence, as well as the overall connectivity $K$ fixed, so it can be understood as a sort of ``micro-canonical ensemble'' \cite{ben2004lect}. Note that this differs from previous studies \cite{goudarzi2012} where the overall network connectivity was allowed to change along the evolutionary dynamics. Our approach permits us to analyze the network performance as a function of network connectivity and, thus, as a function of its dynamical state.
  \end{enumerate}
  
\begin{table}[hbtp]
\begin{centering}
\subfloat[]{\begin{centering}
\begin{tabular}{c|ccc|c}
$$ & $\sigma_{i_2}$ & \cellcolor{red!10} $\overset{\mathbf{\times}}{\sigma_{i_1}}$ & $\sigma_{i_0}$ & $\sigma_{\mathrm{out}}$\tabularnewline
\rowcolor{blue!10}\hline 0 & 0 & 0 & 0 & 0/$\boxempty$\tabularnewline
\rowcolor{blue!10}\hline 1 & 0 & 0 & 1 & 1/$\boxempty$\tabularnewline
\rowcolor{red!10}\hline $\times$ 2 & 0 & 1 & 0 & 1\tabularnewline
\rowcolor{red!10}\hline $\times$ 3 & 0 & 1 & 1 & 0\tabularnewline
\rowcolor{blue!10}\hline 4 & 1 & 0 & 0 & 1/$\boxempty$\tabularnewline
\rowcolor{blue!10}\hline 5 & 1 & 0 & 1 & 0/$\boxempty$\tabularnewline
\rowcolor{red!10}\hline $\times$ 6 & 1 & 1 & 0 & 1\tabularnewline
\rowcolor{red!10}\hline $\times$ 7 & 1 & 1 & 1 & 0\tabularnewline
\end{tabular}
\par\end{centering}
}\subfloat[]{\begin{centering} 
\begin{tabular}{c|ccc|c} $$ & \cellcolor{green!10}$\sigma_{i_2}$ & $\sigma_{i_1}$ & $\sigma_{i_0}$ & $\sigma_\mathrm{out}$\tabularnewline
\hline 0 & 0 & 0 & 0 & 0\tabularnewline
\hline 1 & 0 & 0 & 1 & 1\tabularnewline 
\hline 2 & 0 & 1 & 0 & 1\tabularnewline 
\hline 3 & 0 & 1 & 1 & 0\tabularnewline
\rowcolor{blue!10}\hline 4 & 1 & 0 & 0 & $\boxempty$ \tabularnewline 
\rowcolor{blue!10}\hline 5 & 1 & 0 & 1 & $\boxempty$ \tabularnewline
\rowcolor{blue!10}\hline 6 & 1 & 1 & 0 & $\boxempty$ \tabularnewline
\rowcolor{blue!10}\hline 7 & 1 & 1 & 1 & $\boxempty$ \tabularnewline
\end{tabular} \par\end{centering} } \par\end{centering}  \caption{Examples of the modification of Boolean functions --initially with $3$ inputs and hence $2^3$ possible input configurations-- after the addition or removal of an input node: (a) Link $i_1$ is removed (the connectivity $K_\mathrm{in}$ of the node decreases from $3$ to $2$) the rows $2, 3, 6$, and $7$ (corresponding to $\sigma_{i_1}=1$) are canceled out (marked with $\times$ and redish color); the outputs in rows $0, 1, 4$, and $5$
can be flipped with probability $p=0.25$; (b) Addition of a new link corresponding to input $i_2$ (green color; $K_\mathrm{in}$ of the node increases from $2$ to $3$): outputs for rows $4, 5, 6$, and $7$ are randomly chosen (represented as $\boxempty$). Color code: white cells remain fixed in the mutation, redish ones are removed, and blueish ones involve a random choice.} \end{table}

\subsection*{Assessing network criticality.} We employ the standard method of plotting the Derrida curve in order to determine the dynamical phase of any specific RBN --specified by its topology and the set of its Boolean functions-- and assess how far it operates from criticality.  The method is based in damage spreading dynamics and involves the next steps: (1) take a network $M$ in one specific state, and a copy of it $M'$ in which a single randomly chosen node has changed its state, (2) compute the Hamming distance, $H$, \cite{derrida86,gros2011} between these two networks after one time step ($t=1$; in the asynchronous case nodes are updated following the same random order in both networks), (3) average such a Hamming distance by considering all the possible nodes in the network that can host the initial one-node perturbation, (4) average the previous result over network states. We define the \emph{branching parameter} $B$, as the averaged $H$ after perturbing the different nodes in the network (in some cases, we present results for perturbations only at input/core nodes).  If $B<1$ perturbations shrink on average and the network is said to be subcritical (or in the ordered phase), while if $B>1$ perturbations proliferate and grow on average and the network is supercritical (chaotic or disordered phase).  Finally, in the intermediate case, $B=1$, in which perturbations propagate marginally, the network is critical.

Observe that in networks with some fixed input and output nodes, we can measure $B$ in different ways, depending on whether we flip input nodes or not and on whether we compute the Hamming distance in the whole network or just in the core (excluding input nodes); therefore the concept of criticality might refer to just the core or to the full network.  Finally, in order to determine the critical regime of an ensemble of networks --and not just an individual one-- it is necessary to measure the ensemble average, $\bar{B}$, of $B$.


\begin{thebibliography}{10}
\expandafter\ifx\csname url\endcsname\relax
  \def\url#1{\texttt{#1}}\fi
\expandafter\ifx\csname urlprefix\endcsname\relax\def\urlprefix{URL }\fi
\providecommand{\bibinfo}[2]{#2}
\providecommand{\eprint}[2][]{\url{#2}}

\bibitem{Dogma}
\bibinfo{author}{Crick, F.}
\newblock \bibinfo{title}{Central dogma of molecular biology}.
\newblock \emph{\bibinfo{journal}{Nature}} \textbf{\bibinfo{volume}{227}},
  \bibinfo{pages}{561--563} (\bibinfo{year}{1970}).

\bibitem{buchanan2010}
\bibinfo{author}{Buchanan, M.}
\newblock \emph{\bibinfo{title}{Networks in cell biology}}
  (\bibinfo{publisher}{Cambridge University Press},
  \bibinfo{address}{Cambridge}, \bibinfo{year}{2010}).

\bibitem{kauffman69}
\bibinfo{author}{Kauffman, S.}
\newblock \bibinfo{title}{Metabolic stability and epigenesis in randomly
  constructed genetic nets}.
\newblock \emph{\bibinfo{journal}{J. Theor. Biol.}}
  \textbf{\bibinfo{volume}{22}}, \bibinfo{pages}{437--467}
  (\bibinfo{year}{1969}).

\bibitem{kauffman1993}
\bibinfo{author}{Kauffman, S.~A.}
\newblock \emph{\bibinfo{title}{The origins of order: Self-organization and
  selection in evolution}} (\bibinfo{publisher}{Oxford university press},
  \bibinfo{address}{New York}, \bibinfo{year}{1993}).

\bibitem{gros2011}
\bibinfo{author}{Gros, C.}
\newblock \emph{\bibinfo{title}{Random Boolean networks}}
  (\bibinfo{publisher}{Springer}, \bibinfo{address}{Berlin Heidelberg},
  \bibinfo{year}{2011}).

\bibitem{dejong2002}
\bibinfo{author}{De~Jong, H.}
\newblock \bibinfo{title}{Modeling and simulation of genetic regulatory
  systems: a literature review}.
\newblock \emph{\bibinfo{journal}{J. Comp. Biol.}}
  \textbf{\bibinfo{volume}{9}}, \bibinfo{pages}{67--103}
  (\bibinfo{year}{2002}).

\bibitem{alon2006}
\bibinfo{author}{Alon, U.}
\newblock \emph{\bibinfo{title}{An introduction to systems biology: design
  principles of biological circuits}} (\bibinfo{publisher}{CRC press},
  \bibinfo{address}{London}, \bibinfo{year}{2006}).

\bibitem{droso}
\bibinfo{author}{Albert, R.} \& \bibinfo{author}{Othmer, H.~G.}
\newblock \bibinfo{title}{The topology of the regulatory interactions predicts
  the expression pattern of the segment polarity genes in \textit{Drosophila
  melanogaster}}.
\newblock \emph{\bibinfo{journal}{J. Theor. Biol.}}
  \textbf{\bibinfo{volume}{223}}, \bibinfo{pages}{1--18}
  (\bibinfo{year}{2003}).

\bibitem{yeast}
\bibinfo{author}{Li, F.}, \bibinfo{author}{Long, T.}, \bibinfo{author}{Lu, Y.},
  \bibinfo{author}{Ouyang, Q.} \& \bibinfo{author}{Tang, C.}
\newblock \bibinfo{title}{The yeast cell-cycle network is robustly designed}.
\newblock \emph{\bibinfo{journal}{Proc. Natl. Acad. Sci. USA}}
  \textbf{\bibinfo{volume}{101}}, \bibinfo{pages}{4781--4786}
  (\bibinfo{year}{2004}).

\bibitem{bornholdt}
\bibinfo{author}{Bornholdt, S.}
\newblock \bibinfo{title}{Less is more in modeling large genetic networks}.
\newblock \emph{\bibinfo{journal}{Science}} \textbf{\bibinfo{volume}{310}},
  \bibinfo{pages}{449} (\bibinfo{year}{2005}).

\bibitem{drossel2008}
\bibinfo{author}{Drossel, B.}
\newblock \bibinfo{title}{Random boolean networks}.
\newblock In \bibinfo{editor}{Schuster, H.~G.} (ed.)
  \emph{\bibinfo{booktitle}{Reviews of nonlinear dynamics and complexity}},
  vol.~\bibinfo{volume}{1}, chap.~\bibinfo{chapter}{3},
  \bibinfo{pages}{69--110} (\bibinfo{publisher}{Wiley VCH},
  \bibinfo{address}{Weinheim}, \bibinfo{year}{2008}).

\bibitem{aldana-review}
\bibinfo{author}{Aldana, M.}, \bibinfo{author}{Coppersmith, S.} \&
  \bibinfo{author}{Kadanoff, L.~P.}
\newblock \bibinfo{title}{Boolean dynamics with random couplings}.
\newblock In \bibinfo{editor}{Kaplan, E.}, \bibinfo{editor}{Marsden, J.~E.} \&
  \bibinfo{editor}{Sreenivasan, K.~R.} (eds.)
  \emph{\bibinfo{booktitle}{Perspectives and Problems in Nonlinear Science}},
  \bibinfo{pages}{23--89} (\bibinfo{publisher}{Springer-Verlag},
  \bibinfo{address}{New York}, \bibinfo{year}{2003}).

\bibitem{derrida86}
\bibinfo{author}{Derrida, B.} \& \bibinfo{author}{Pomeau, Y.}
\newblock \bibinfo{title}{Random networks of automata: a simple annealed
  approximation}.
\newblock \emph{\bibinfo{journal}{Europhys. Lett.}}
  \textbf{\bibinfo{volume}{1}}, \bibinfo{pages}{45} (\bibinfo{year}{1986}).

\bibitem{kauffman2003}
\bibinfo{author}{Kauffman, S.}, \bibinfo{author}{Peterson, C.},
  \bibinfo{author}{Samuelsson, B.} \& \bibinfo{author}{Troein, C.}
\newblock \bibinfo{title}{Random boolean network models and the yeast
  transcriptional network}.
\newblock \emph{\bibinfo{journal}{Proc. Natl. Acad. Sci. USA}}
  \textbf{\bibinfo{volume}{100}}, \bibinfo{pages}{14796--14799}
  (\bibinfo{year}{2003}).

\bibitem{langton1990}
\bibinfo{author}{Langton, C.~G.}
\newblock \bibinfo{title}{Computation at the edge of chaos: phase transitions
  and emergent computation}.
\newblock \emph{\bibinfo{journal}{Physica D}} \textbf{\bibinfo{volume}{42}},
  \bibinfo{pages}{12--37} (\bibinfo{year}{1990}).

\bibitem{maas2002}
\bibinfo{author}{Maass, W.}, \bibinfo{author}{Natschl{\"a}ger, T.} \&
  \bibinfo{author}{Markram, H.}
\newblock \bibinfo{title}{Real-time computing without stable states: A new
  framework for neural computation based on perturbations}.
\newblock \emph{\bibinfo{journal}{Neural Comput.}}
  \textbf{\bibinfo{volume}{14}}, \bibinfo{pages}{2531--2560}
  (\bibinfo{year}{2002}).

\bibitem{ber-Nat}
\bibinfo{author}{Bertschinger, N.} \& \bibinfo{author}{Natschlager, T.}
\newblock \bibinfo{title}{Real-time computation at the edge of chaos in
  recurrent neural networks}.
\newblock \emph{\bibinfo{journal}{Neural Comput.}}
  \textbf{\bibinfo{volume}{16}}, \bibinfo{pages}{1413--1436}
  (\bibinfo{year}{2004}).

\bibitem{aldana2007}
\bibinfo{author}{Aldana, M.}, \bibinfo{author}{Balleza, E.},
  \bibinfo{author}{Kauffman, S.} \& \bibinfo{author}{Resendiz, O.}
\newblock \bibinfo{title}{Robustness and evolvability in genetic regulatory
  networks}.
\newblock \emph{\bibinfo{journal}{J. Theor. Biol.}}
  \textbf{\bibinfo{volume}{245}}, \bibinfo{pages}{433--448}
  (\bibinfo{year}{2007}).

\bibitem{Aldana2012}
\bibinfo{author}{Sosa, C.~T.}, \bibinfo{author}{Huang, S.} \&
  \bibinfo{author}{Aldana, M.}
\newblock \bibinfo{title}{Criticality is an emergent property of genetic
  networks that exhibit evolvability}.
\newblock \emph{\bibinfo{journal}{PLoS Comp. Biol.}}
  \textbf{\bibinfo{volume}{8}}, \bibinfo{pages}{e1002669}
  (\bibinfo{year}{2012}).

\bibitem{ribeiro2008}
\bibinfo{author}{Ribeiro, A.~S.}, \bibinfo{author}{Kauffman, S.~A.},
  \bibinfo{author}{Lloyd-Price, J.}, \bibinfo{author}{Samuelsson, B.} \&
  \bibinfo{author}{Socolar, J.~E.}
\newblock \bibinfo{title}{Mutual information in random boolean models of
  regulatory networks}.
\newblock \emph{\bibinfo{journal}{Phys. Rev. E}} \textbf{\bibinfo{volume}{77}},
  \bibinfo{pages}{011901} (\bibinfo{year}{2008}).

\bibitem{basin}
\bibinfo{author}{Krawitz, P.} \& \bibinfo{author}{Shmulevich, I.}
\newblock \bibinfo{title}{Basin entropy in boolean network ensembles}.
\newblock \emph{\bibinfo{journal}{Phys. Rev. Lett.}}
  \textbf{\bibinfo{volume}{98}}, \bibinfo{pages}{158701}
  (\bibinfo{year}{2007}).

\bibitem{torres2012}
\bibinfo{author}{Torres-Sosa, C.}, \bibinfo{author}{Huang, S.} \&
  \bibinfo{author}{Aldana, M.}
\newblock \bibinfo{title}{Criticality is an emergent property of genetic
  networks that exhibit evolvability}.
\newblock \emph{\bibinfo{journal}{PLoS Comp. Biol.}}
  \textbf{\bibinfo{volume}{8}}, \bibinfo{pages}{e1002669}
  (\bibinfo{year}{2012}).

\bibitem{Plenz-Functional}
\bibinfo{author}{Shew, W.~L.} \& \bibinfo{author}{Plenz, D.}
\newblock \bibinfo{title}{The functional benefits of criticality in the
  cortex}.
\newblock \emph{\bibinfo{journal}{Neuroscientist}}
  \textbf{\bibinfo{volume}{19}}, \bibinfo{pages}{88--100}
  (\bibinfo{year}{2013}).

\bibitem{Kinouchi-Copelli}
\bibinfo{author}{Kinouchi, O.} \& \bibinfo{author}{Copelli, M.}
\newblock \bibinfo{title}{Optimal dynamical range of excitable networks at
  criticality}.
\newblock \emph{\bibinfo{journal}{Nat. Phys.}} \textbf{\bibinfo{volume}{2}},
  \bibinfo{pages}{348--351} (\bibinfo{year}{2006}).

\bibitem{Kaneko2012}
\bibinfo{author}{Furusawa, C.} \& \bibinfo{author}{Kaneko, K.}
\newblock \bibinfo{title}{Adaptation to optimal cell growth through
  self-organized criticality.}
\newblock \emph{\bibinfo{journal}{Phys. Rev. Lett.}}
  \textbf{\bibinfo{volume}{108}}, \bibinfo{pages}{208103}
  (\bibinfo{year}{2012}).

\bibitem{Physics}
\bibinfo{author}{Chat{\'e}, H.} \& \bibinfo{author}{Mu{\~n}oz, M.}
\newblock \bibinfo{title}{Insect swarms go critical}.
\newblock \emph{\bibinfo{journal}{Physics}} \textbf{\bibinfo{volume}{7}},
  \bibinfo{pages}{120} (\bibinfo{year}{2014}).

\bibitem{balleza2008}
\bibinfo{author}{Balleza, E.} \emph{et~al.}
\newblock \bibinfo{title}{Critical dynamics in genetic regulatory networks:
  examples from four kingdoms}.
\newblock \emph{\bibinfo{journal}{PLoS One}} \textbf{\bibinfo{volume}{3}},
  \bibinfo{pages}{e2456} (\bibinfo{year}{2008}).

\bibitem{nykter2008}
\bibinfo{author}{Nykter, M.} \emph{et~al.}
\newblock \bibinfo{title}{Gene expression dynamics in the macrophage exhibit
  criticality}.
\newblock \emph{\bibinfo{journal}{Proc. Natl. Acad. Sci. USA}}
  \textbf{\bibinfo{volume}{105}}, \bibinfo{pages}{1897--1900}
  (\bibinfo{year}{2008}).

\bibitem{shmulevich2005}
\bibinfo{author}{Shmulevich, I.}, \bibinfo{author}{Kauffman, S.~A.} \&
  \bibinfo{author}{Aldana, M.}
\newblock \bibinfo{title}{Eukaryotic cells are dynamically ordered or critical
  but not chaotic}.
\newblock \emph{\bibinfo{journal}{Proc. Natl. Acad. Sci. USA}}
  \textbf{\bibinfo{volume}{102}}, \bibinfo{pages}{13439--13444}
  (\bibinfo{year}{2005}).

\bibitem{hidalgo2014}
\bibinfo{author}{Hidalgo, J.} \emph{et~al.}
\newblock \bibinfo{title}{Information-based fitness and the emergence of
  criticality in living systems}.
\newblock \emph{\bibinfo{journal}{Proc. Natl. Acad. Sci. USA}}
  \textbf{\bibinfo{volume}{111}}, \bibinfo{pages}{10095--10100}
  (\bibinfo{year}{2014}).

\bibitem{goudarzi2012}
\bibinfo{author}{Goudarzi, A.}, \bibinfo{author}{Teuscher, C.},
  \bibinfo{author}{Gulbahce, N.} \& \bibinfo{author}{Rohlf, T.}
\newblock \bibinfo{title}{Emergent criticality through adaptive information
  processing in boolean networks}.
\newblock \emph{\bibinfo{journal}{Phys. Rev. Lett.}}
  \textbf{\bibinfo{volume}{108}}, \bibinfo{pages}{128702}
  (\bibinfo{year}{2012}).

\bibitem{gupta}
\bibinfo{author}{Guptasarma, P.}
\newblock \bibinfo{title}{Does replication-induced transcription regulate
  synthesis of the myriad low copy number proteins of \textit{Escherichia coli}?}
\newblock \emph{\bibinfo{journal}{Bioessays}} \textbf{\bibinfo{volume}{17}},
  \bibinfo{pages}{987--997} (\bibinfo{year}{1995}).

\bibitem{noise2000}
\bibinfo{author}{Schwikowski, B.}, \bibinfo{author}{Uetz, P.} \&
  \bibinfo{author}{Fields, S.}
\newblock \bibinfo{title}{A network of protein--protein interactions in yeast}.
\newblock \emph{\bibinfo{journal}{Nat. Biotechnol.}}
  \textbf{\bibinfo{volume}{18}}, \bibinfo{pages}{1257--1261}
  (\bibinfo{year}{2000}).

\bibitem{elowitz2002}
\bibinfo{author}{Elowitz, M.~B.}, \bibinfo{author}{Levine, A.~J.},
  \bibinfo{author}{Siggia, E.~D.} \& \bibinfo{author}{Swain, P.~S.}
\newblock \bibinfo{title}{Stochastic gene expression in a single cell}.
\newblock \emph{\bibinfo{journal}{Science}} \textbf{\bibinfo{volume}{297}},
  \bibinfo{pages}{1183--1186} (\bibinfo{year}{2002}).

\bibitem{elowitz2010}
\bibinfo{author}{Eldar, A.} \& \bibinfo{author}{Elowitz, M.~B.}
\newblock \bibinfo{title}{Functional roles for noise in genetic circuits}.
\newblock \emph{\bibinfo{journal}{Nature}} \textbf{\bibinfo{volume}{467}},
  \bibinfo{pages}{167--173} (\bibinfo{year}{2010}).

\bibitem{losick2008}
\bibinfo{author}{Losick, R.} \& \bibinfo{author}{Desplan, C.}
\newblock \bibinfo{title}{Stochasticity and cell fate}.
\newblock \emph{\bibinfo{journal}{Science}} \textbf{\bibinfo{volume}{320}},
  \bibinfo{pages}{65--68} (\bibinfo{year}{2008}).

\bibitem{balazsi2011}
\bibinfo{author}{Bal{\'a}zsi, G.}, \bibinfo{author}{van Oudenaarden, A.} \&
  \bibinfo{author}{Collins, J.~J.}
\newblock \bibinfo{title}{Cellular decision making and biological noise: from
  microbes to mammals}.
\newblock \emph{\bibinfo{journal}{Cell}} \textbf{\bibinfo{volume}{144}},
  \bibinfo{pages}{910--925} (\bibinfo{year}{2011}).

\bibitem{tkacik}
\bibinfo{author}{Tka{\v{c}}ik, G.} \& \bibinfo{author}{Walczak, A.~M.}
\newblock \bibinfo{title}{Information transmission in genetic regulatory
  networks: a review}.
\newblock \emph{\bibinfo{journal}{J. Phys. Condens. Mat.}}
  \textbf{\bibinfo{volume}{23}}, \bibinfo{pages}{153102}
  (\bibinfo{year}{2011}).

\bibitem{stern}
\bibinfo{author}{Stern, M.~D.}
\newblock \bibinfo{title}{Emergence of homeostasis and ``noise imprinting'' in
  an evolution model}.
\newblock \emph{\bibinfo{journal}{Proc. Natl. Acad. Sci. USA}}
  \textbf{\bibinfo{volume}{96}}, \bibinfo{pages}{10746--10751}
  (\bibinfo{year}{1999}).

\bibitem{darabos}
\bibinfo{author}{Darabos, C.}, \bibinfo{author}{Tomassini, M.} \&
  \bibinfo{author}{Giacobini, M.}
\newblock \bibinfo{title}{Dynamics of unperturbed and noisy generalized boolean
  networks}.
\newblock \emph{\bibinfo{journal}{J. Theor. Biol.}}
  \textbf{\bibinfo{volume}{260}}, \bibinfo{pages}{531--544}
  (\bibinfo{year}{2009}).

\bibitem{peixoto2012}
\bibinfo{author}{Peixoto, T.~P.}
\newblock \bibinfo{title}{Emergence of robustness against noise: A structural
  phase transition in evolved models of gene regulatory networks}.
\newblock \emph{\bibinfo{journal}{Phys. Rev. E}} \textbf{\bibinfo{volume}{85}},
  \bibinfo{pages}{041908} (\bibinfo{year}{2012}).

\bibitem{gershenson}
\bibinfo{author}{Gershenson, C.}
\newblock \bibinfo{title}{Classification of random boolean networks}.
\newblock In \bibinfo{editor}{Standish, R.~K.}, \bibinfo{editor}{Bedau, M.~A.}
  \& \bibinfo{editor}{Abbas, H.~A.} (eds.)
  \emph{\bibinfo{booktitle}{Proceedings of the eighth international conference
  on Artificial life}}, \bibinfo{pages}{1--8} (\bibinfo{publisher}{The MIT
  Press}, \bibinfo{address}{Cambridge}, \bibinfo{year}{2002}).

\bibitem{drossel2005}
\bibinfo{author}{Greil, F.} \& \bibinfo{author}{Drossel, B.}
\newblock \bibinfo{title}{Dynamics of critical kauffman networks under
  asynchronous stochastic update}.
\newblock \emph{\bibinfo{journal}{Phys. Rev. Lett.}}
  \textbf{\bibinfo{volume}{95}}, \bibinfo{pages}{048701}
  (\bibinfo{year}{2005}).

\bibitem{peixoto-drossel2010}
\bibinfo{author}{Schmal, C.}, \bibinfo{author}{Peixoto, T.~P.} \&
  \bibinfo{author}{Drossel, B.}
\newblock \bibinfo{title}{Boolean networks with robust and reliable
  trajectories}.
\newblock \emph{\bibinfo{journal}{New J. Phys.}} \textbf{\bibinfo{volume}{12}},
  \bibinfo{pages}{113054} (\bibinfo{year}{2010}).

\bibitem{Wolfram2002}
\bibinfo{author}{Wolfram, S.}
\newblock \emph{\bibinfo{title}{A new kind of science}},
  vol.~\bibinfo{volume}{5} (\bibinfo{publisher}{Wolfram media},
  \bibinfo{address}{Champaign}, \bibinfo{year}{2002}).

\bibitem{stark2006}
\bibinfo{author}{Stark, C.} \emph{et~al.}
\newblock \bibinfo{title}{Biogrid: a general repository for interaction
  datasets}.
\newblock \emph{\bibinfo{journal}{Nucleic Acids Res.}}
  \textbf{\bibinfo{volume}{34}}, \bibinfo{pages}{D535--D539}
  (\bibinfo{year}{2006}).

\bibitem{feist2009}
\bibinfo{author}{Feist, A.~M.}, \bibinfo{author}{Herrg{\aa}rd, M.~J.},
  \bibinfo{author}{Thiele, I.}, \bibinfo{author}{Reed, J.~L.} \&
  \bibinfo{author}{Palsson, B.~{\O}.}
\newblock \bibinfo{title}{Reconstruction of biochemical networks in
  microorganisms}.
\newblock \emph{\bibinfo{journal}{Nat. Rev. Microbiol.}}
  \textbf{\bibinfo{volume}{7}}, \bibinfo{pages}{129--143}
  (\bibinfo{year}{2009}).

\bibitem{chang2011}
\bibinfo{author}{Chang, R.~L.} \emph{et~al.}
\newblock \bibinfo{title}{Metabolic network reconstruction of chlamydomonas
  offers insight into light-driven algal metabolism}.
\newblock \emph{\bibinfo{journal}{Mol. Sys. Biol.}}
  \textbf{\bibinfo{volume}{7}}, \bibinfo{pages}{518} (\bibinfo{year}{2011}).

\bibitem{oh2007}
\bibinfo{author}{Oh, Y.-K.}, \bibinfo{author}{Palsson, B.~O.},
  \bibinfo{author}{Park, S.~M.}, \bibinfo{author}{Schilling, C.~H.} \&
  \bibinfo{author}{Mahadevan, R.}
\newblock \bibinfo{title}{Genome-scale reconstruction of metabolic network in
  \textit{Bacillus subtilis} based on high-throughput phenotyping and gene essentiality
  data}.
\newblock \emph{\bibinfo{journal}{J. Biol. Chem.}}
  \textbf{\bibinfo{volume}{282}}, \bibinfo{pages}{28791--28799}
  (\bibinfo{year}{2007}).

\bibitem{balazsi2005}
\bibinfo{author}{Bal\'{a}zsi, G.}, \bibinfo{author}{Barab\'{a}si, A.-L.} \&
  \bibinfo{author}{Oltvai, Z.~N.}
\newblock \bibinfo{title}{Topological units of environmental signal processing
  in the transcriptional regulatory network of \textit{Escherichia coli}}.
\newblock \emph{\bibinfo{journal}{Proc. Natl. Acad. Sci. USA}}
  \textbf{\bibinfo{volume}{102}}, \bibinfo{pages}{7841--7846}
  (\bibinfo{year}{2005}).

\bibitem{ma2007}
\bibinfo{author}{Ma, S.}, \bibinfo{author}{Gong, Q.} \&
  \bibinfo{author}{Bohnert, H.~J.}
\newblock \bibinfo{title}{An arabidopsis gene network based on the graphical
  gaussian model}.
\newblock \emph{\bibinfo{journal}{Genome Res.}} \textbf{\bibinfo{volume}{17}},
  \bibinfo{pages}{1614--1625} (\bibinfo{year}{2007}).

\bibitem{balazsi2008}
\bibinfo{author}{Bal\'{a}zsi, G.}, \bibinfo{author}{Heath, A.~P.},
  \bibinfo{author}{Shi, L.} \& \bibinfo{author}{Gennaro, M.~L.}
\newblock \bibinfo{title}{The temporal response of the \textit{Mycobacterium
  tuberculosis} gene regulatory network during growth arrest}.
\newblock \emph{\bibinfo{journal}{Mol. Sys. Biol.}}
  \textbf{\bibinfo{volume}{4}} (\bibinfo{year}{2008}).

\bibitem{sanz2011trans}
\bibinfo{author}{Sanz, J.} \emph{et~al.}
\newblock \bibinfo{title}{The transcriptional regulatory network of
  \textit{Mycobacterium tuberculosis}}.
\newblock \emph{\bibinfo{journal}{PLoS One}} \textbf{\bibinfo{volume}{6}},
  \bibinfo{pages}{e22178} (\bibinfo{year}{2011}).

\bibitem{galan2011}
\bibinfo{author}{Gal\'{a}n-V\'{a}squez, E.}, \bibinfo{author}{Luna, B.} \&
  \bibinfo{author}{Mart\'{\i}nez-Antonio, A.}
\newblock \bibinfo{title}{The regulatory network of \textit{Pseudomonas aeruginosa}}.
\newblock \emph{\bibinfo{journal}{Microb. Inform. Exp.}}
  \textbf{\bibinfo{volume}{1}}, \bibinfo{pages}{1--11} (\bibinfo{year}{2011}).

\bibitem{guelzim2002}
\bibinfo{author}{Guelzim, N.}, \bibinfo{author}{Bottani, S.},
  \bibinfo{author}{Bourgine, P.} \& \bibinfo{author}{K\'{e}p\`{e}s, F.}
\newblock \bibinfo{title}{Topological and causal structure of the yeast
  transcriptional regulatory network}.
\newblock \emph{\bibinfo{journal}{Nat. Genet.}} \textbf{\bibinfo{volume}{31}},
  \bibinfo{pages}{60--63} (\bibinfo{year}{2002}).

\bibitem{leclerc2008}
\bibinfo{author}{Leclerc, R.~D.}
\newblock \bibinfo{title}{Survival of the sparsest: robust gene networks are
  parsimonious}.
\newblock \emph{\bibinfo{journal}{Mol. Sys. Biol.}}
  \textbf{\bibinfo{volume}{4}} (\bibinfo{year}{2008}).

\bibitem{drossel2009}
\bibinfo{author}{Drossel, B.} \& \bibinfo{author}{Greil, F.}
\newblock \bibinfo{title}{Critical boolean networks with scale-free in-degree
  distribution}.
\newblock \emph{\bibinfo{journal}{Phys. Rev. E}} \textbf{\bibinfo{volume}{80}},
  \bibinfo{pages}{026102} (\bibinfo{year}{2009}).

\bibitem{drossel2010}
\bibinfo{author}{Szejka, A.} \& \bibinfo{author}{Drossel, B.}
\newblock \bibinfo{title}{Evolution of boolean networks under selection for a
  robust response to external inputs yields an extensive neutral space}.
\newblock \emph{\bibinfo{journal}{Phys. Rev. E}} \textbf{\bibinfo{volume}{81}},
  \bibinfo{pages}{021908} (\bibinfo{year}{2010}).

\bibitem{aldana2003}
\bibinfo{author}{Aldana, M.}
\newblock \bibinfo{title}{Boolean dynamics of networks with scale-free
  topology}.
\newblock \emph{\bibinfo{journal}{Physica D}} \textbf{\bibinfo{volume}{185}},
  \bibinfo{pages}{45--66} (\bibinfo{year}{2003}).

\bibitem{ben2004lect}
\bibinfo{author}{Ben-Naim, E.}, \bibinfo{author}{Frauenfelder, H.} \&
  \bibinfo{author}{Toroczkai, Z.}
\newblock \bibinfo{title}{Complex networks}.
\newblock In \emph{\bibinfo{booktitle}{Lecture Notes in Physics}}, \textbf{\bibinfo{volume}{650}}, (\bibinfo{publisher}{Springer},
  \bibinfo{address}{Berlin Heidelberg}, \bibinfo{year}{2004}).

\end{thebibliography}

\section*{Acknowledgments}
 We acknowledge the Spanish-MINECO grant FIS2013-43201-P (FEDER
  funds) for financial support. We thank P. Moretti for very useful comments.
  
\section*{Author Contributions}
J.H. and M.A.M. conceived the project, P.V. and J.M.R. performed the numerical simulations and prepared the figures. P.V. J.H. and M.A.M. wrote the main manuscript text. All authors reviewed the manuscript.

\section*{Additional Information}
The authors declare no competing financial interests.

 \end{document}